# MAE-GAN: A Novel Strategy for Simultaneous Super-resolution Reconstruction and Denoising of Post-stack Seismic Profile

Wenshuo Yu, Shiqi Dong, Shaoping Lu and Xintong Dong

*Abstract*—Post-stack seismic profiles are images reflecting containing geological structures which provides a critical foundation for understanding the distribution of oil and gas resources. However, due to the limitations of seismic acquisition equipment and data collecting geometry, the post-stack profiles suffer from low resolution and strong noise issues, which severely affects subsequent seismic interpretation. To better enhance the spatial resolution and signal-to-noise ratio of post-seismic profiles, a multi-scale attention encoder-decoder network based on generative adversarial network (MAE-GAN) is proposed. This method improves the resolution of post-stack profiles, and effectively suppresses noises and recovers weak signals as well. A multi-scale residual module is proposed to extract geological features under different receptive fields. At the same time, an attention module is designed to further guide the network to focus on important feature information. Additionally, to better recover the global and local information of post-stack profiles, an adversarial network based on a Markov discriminator is proposed. Finally, by introducing an edge information preservation loss function, the conventional loss function of the Generative Adversarial Network is improved, which enables better recovery of the edge information of the original post-stack profiles. Experimental results on simulated and field post-stack profiles demonstrate that the proposed MAE-GAN method outperforms two advanced convolutional neural network-based methods in noise suppression and weak signal recovery. Furthermore, the profiles reconstructed by the MAE-GAN method preserve more geological structures.

*Index Terms*—Generative adversarial network (GAN), Post-stack seismic profile, super-resolution reconstruction, seismic noise suppression, dual attention mechanism, multi-scale residual model.

## I. INTRODUCTION

Due to the limitations of seismic acquisition and processing technologies, field post-stack seismic data often suffer from low resolution problem and noise interference, posing challenges to subsequent seismic interpretation [1], [2]. Therefore, it is crucial to enhance the resolution and signal-to-noise ratio of post-stack profiles in order to improve the accuracy and reliability of seismic interpretation in oil and gas exploration.

The objective of super-resolution reconstruction of post-stack seismic profiles is to enhance the resolution and signal-to-noise ratio of the profiles, while broadening the frequency band, especially improving the main frequency of the signal, and restoring the amplitude of weak signals. Existing post-stack seismic profile super-resolution reconstruction methods are primarily divided into four types: deconvolution-based methods [3], [4], absorption compensation-based methods [5], [6], spread spectrum-based methods [7], [8], and deep learning-based methods. Deconvolution methods compress seismic waveforms and extend their frequency spectrum to recover the original signal, enhancing resolution [9], [10]. However, they can amplify noises, leading to distortion. Absorption compensation-based methods restore and enhance high-frequency components of seismic signals, improving spatial resolution [11], [12]. For example, Braga et al. [13] proposed an inverse Q filtering method based on continuous wavelet transform (CWT) to correct amplitude and phase distortions. However, these methods rely on accurate medium absorption models and have complex processing procedures. Spread spectrum-based methods use time-frequency decomposition to process high- and low-frequency components, extending the frequency band range and enhancing resolution. Alaei et al. [14] improved the vertical resolution of seismic images by combining Gabor deconvolution and wavelet scaling techniques. However, these methods can over-amplify noise or non-geological signals, leading to artifacts in the reconstructed profile.

In recent years, with the continuous improvement of hardware computational capabilities, deep learning technology has developed rapidly. Deep learning can process complex data patterns and features, enhancing the accuracy of analysis

This work was supported in part by the National Natural Science Foundation of China under grants 42204114 and 42074123, in part by the Scientific and Technological Developing Scheme Project of Jilin Province under Grant 20220203030SF, and in part by the Natural Science Foundation of Jilin Province under Grant 20230101073JC. (*Corresponding author: Xintong Dong*)

Wenshuo Yu and Xintong Dong are with the College of Instrumentation and Electrical Engineering, Jilin University, Changchun, Jilin 130026, China, (e-mail: yuwenshuo1998@163.com; 18186829038@163.com).

Shiqi Dong is with the Key Laboratory of Modern Power System Simulation and the Control and Renewable Energy Technology, Ministry of Education, Jilin City, Jilin 132012, China, and also with the Department of Communication Engineering, College of Electric Engineering, Northeast Electric Power University, Jilin City, Jilin 132012, China (e-mail: dsq1994@126.com).

Shaoping Lu is with the School of Earth Sciences and Engineering, Sun Yat-Sen University, Guangzhou, Guangdong 510275, China, the Southern Marine Science and Engineering Guangdong Laboratory (Zhuhai), Zhuhai, Guangdong 519000, China, and the Guangdong Provincial Key Lab of Geodynamics and Geohazards, Sun Yat-Sen University, Guangzhou, Guangdong 510275, China. (e-mail:lushaoping@mail.sysu.edu.cn)

and prediction, thus gradually replacing traditional signal processing methods [15], [16]. The post-stack seismic profile super-resolution reconstruction method based on deep learning involves training deep neural networks, such as convolutional neural networks (CNNs) or generative adversarial networks (GANs), to automatically learn the nonlinear mapping relationship between low-resolution seismic data and high-resolution seismic data. Li et al. [17] employed a loss function that combines L1 loss and multiscale structural similarity loss to enhance the perceptual quality of seismic images, thus better improving geological structures and stratigraphic features. Gao et al. [18] used a U-shaped network that incorporates residual block and attention mechanism to learn the mapping relationship between low and high resolution. Additionally, they employed a hybrid loss function combining L1 loss and structural similarity loss to optimize the network, improving its ability to recognize geometric features characterized by structural amplitude changes. Wang et al. [19] proposed an unsupervised deep learning method to learns features from seismic data by integrating physical constraints and prior knowledge and then uses these features to reconstruct high-resolution data. Hamida et al. [20] propose a loss function that incorporates facies information to train super-resolution models, thereby improving the resolution of seismic images while enhancing the accuracy of facies classification. Zeng et al. [21] improved the clarity of seismic images by integrating multi-scale convolutional kernels, a dual-branch network, and an enhanced reconstruction module. Zhang et al. [22] improved the performance of seismic data resolution enhancement by introducing ground data features through linear operations. Choi et al. [23] proposed a spectral enhancement method considering features of seismic field data, trained using a convolutional U-Net model and incorporating numerous synthetic data and prior information, improves the resolution of seismic data. Zhou et al. [24] proposed a deep learning framework based on geophysical priors, which improves the quality of seismic data by incorporating fault priors to guide model training.

GANs are a specialized type of deep learning model, divided into two components: the generative network and the discriminative network. The performance of the model is enhanced by the adversarial game played between these two networks [25], [26]. GANs have been extensively used in fields such as denoising [27], [28], style transfer [29], and super-resolution reconstruction [30], [31], as well as in other seismic applications including noise suppression [32], [33], interpolation [34], [35], and inversion [36], [37]. Recently, researchers have applied GANs to the task of post-stack seismic profile super-resolution reconstruction. Lin et al. [38] optimized the GAN model by incorporating VGG loss into the loss function of the original GAN, thereby enhancing the clarity and detail of seismic images. Sun et al. [39] used residual learning and back-projection units to achieve random noise suppression and super-resolution reconstruction of seismic profiles.

Although post-stack seismic profile super-resolution reconstruction methods based on GANs have more effective super-resolution performance compared to traditional deep learning models, they often struggle to recover weak signals, resulting in unclear geological structures such as faults. Therefore, we propose a multi-scale attention encoder-decoder network based on generative adversarial network (MAE-GAN) for simultaneous post-stack seismic profile super-resolution reconstruction and denoising to improve the perceptual quality of the post-stack profiles.

The main contributions of MAE-GAN are summarised as follows:

1. For the generative network, a multi-scale residual module is designed to extract more features of post-stack seismic profiles from different scales of receptive fields and to prevent vanishing gradients during the training process. Simultaneously, a dual attention module is proposed to allocate greater weight to important feature information, allowing the generative network to extract critical features more effectively. Additionally, Resnet-D is introduced to replace pooling layers, addressing the loss of feature information caused by pooling operations through decoupled downsampling.

2. For the discriminator network, an adversarial network based on a Markovian discriminator is designed to replace the traditional fully connected layer adversarial network, which reduces the computational cost of the network and enhances the discriminative ability by discriminating input images in patches, thereby aiding in improving the super-resolution reconstruction and noise suppression performance of the generative network.

3. To better restore the geological structure of post-stack profiles, an improved loss function is proposed by incorporating an edge information preservation loss function into the conventional loss function of GAN. The edge information preservation loss function more accurately measures the differences in edge information features between the reconstructed seismic profile and the original high-resolution seismic profile.

Both synthetic and real post-stack seismic profiles are used to validate the effectiveness of MAE-GAN. Experimental results demonstrate that, compared to other advanced super-resolution reconstruction methods, the proposed MAE-GAN method can effectively recover weak signals while enhancing the resolution of post-stack seismic profiles to make the geological structures clearer.

## II. Description Of Network Architecture

In this section, we describe the architecture of the proposed MAE-GAN and its main network components. MAE-GAN consists of a generative network and an adversarial network. The generative network is used to simultaneously enhance the resolution of post-stack seismic profiles and suppress noise. The adversarial network is used to discriminate whether the input image is a seismic profile reconstructed by the generator or the original high-resolution seismic profile.

*A. The architecture of Generative Network in MAE-GAN*

The proposed MAE-GAN is shown in Fig. 1. The generative network consists of an encoder and a decoder. The encoder reduces the dimensions of the feature maps by successive downsampling operations, which facilitates the extraction of low-frequency information of the seismic profiles. The decoder reconstructs high-resolution seismic profiles by upsampling operations, thereby better extracting and restoring the details and high-frequency information of the seismic profiles. Additionally, we introduce skip connections to better fuse low-frequency and high-frequency information, thus preventing information loss. Next, we will first introduce the basic components of the generative network, and then provide a comprehensive overview of the complete generative network.

In the proposed generative network, in order to extract feature information of post-stack seismic profiles more effectively, we initially designed a multi-scale residual module as shown in Fig. 2. This multi-scale residual module consists of two branches. The first branch is composed of a skip connection with a 1×1 convolutional layer, which preserves the initial information of the input feature map and adjusts the number of channel. The second branch consists of three parallel feature extraction groups with different receptive fields, an independent LeakyReLU activation function, and a 3×3 convolutional layer with LeakyReLU activation function. Each feature extraction group comprises a dilated convolution layer and two 3×3 convolutional layers, with the dilation rates of the dilated convolutions in the three parallel groups being 1, 2, and 3, respectively. This allows for effective feature extraction under different receptive fields. The output feature maps of the three parallel feature extraction groups are fused by the element-wise sum, aggregating feature information from different scales. The fused output feature map then serves as the input to an independent LeakyReLU activation function and a 3×3 convolutional layer with LeakyReLU activation function. The LeakyReLU activation function introduces non-linear characteristics and prevents the gradient vanishing problem during training, while the 3×3 convolutional layer extracts more effective deep feature information. Finally, the output feature maps of both branches are fused by the element-wise sum to obtain the output feature map of the multi-scale residual module.

In the proposed generative network, we have also designed a dual attention module as shown in Fig. 3. This module enhances the network's ability to extract significant features by assigning greater weights to important features, thereby mitigating the interference of noises. The dual attention module consists of a channel attention module and a spatial attention module, which are cascaded together. The channel attention module comprises two branches with the same input feature map. The first branch is a skip connection designed to prevent information loss. The second branch consists of a global average pooling layer, a 1×1 convolutional layer, and a sigmoid activation function. The global average pooling layer reduces the size of feature map to 1×1, the 1×1 convolutional layer is used for extracting channel feature information, and the sigmoid activation function is utilized to obtain channel attention weights. The feature map output of the first branch is multiplied element-wise with the channel attention weights output of the second branch to obtain the channel attention feature map. This feature map serves as the input for the spatial attention module. The spatial attention module includes two branches, where the first is a skip connection to prevent loss of feature information, and the second comprises three smaller sub-branches, a 1×1 convolutional layer, a 3×3 convolutional layer, and a sigmoid activation function. The first sub-branch is a channel global average pooling layer that aggregates global geological structure information, the second sub-branch is a channel global max pooling layer for aggregating detailed geological structure information, and the third sub-branch consists of a downsampling layer, three cascaded 3×3 convolutional layers, and an upsampling layer. The outputs of these three sub-branches are merged by channel concatenation to integrate features, and the merged output is further processed by the 1×1 convolutional layer and the 3×3 convolutional layer to adjust channel numbers and extract deeper spatial feature information. The deep feature-extracted output map is then processed by the sigmoid activation function to obtain spatial attention weights. The feature map output of the first branch is multiplied element-wise with the spatial attention weights output of the second branch to produce the output feature map of the dual attention module. The computational process of the dual attention module can be expressed as:

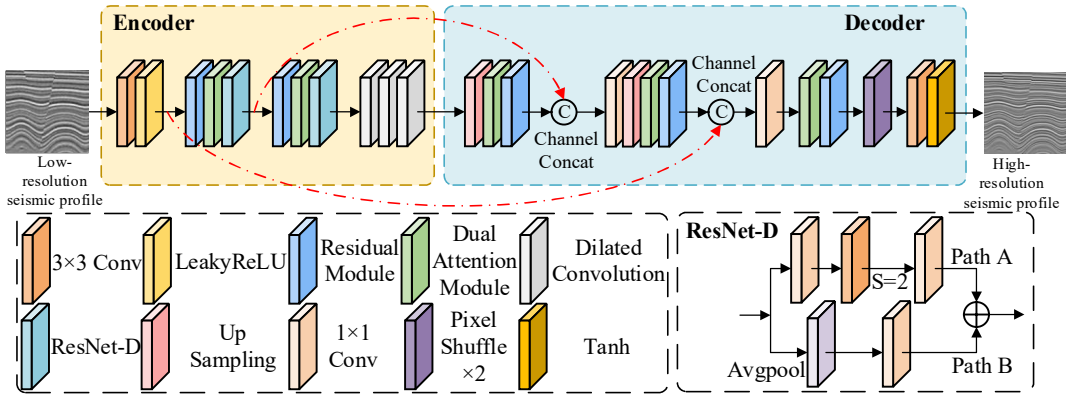

**Fig. 1.** The generative network of the proposed MAE-GAN. (The generative network consists of an encoder and a decoder)

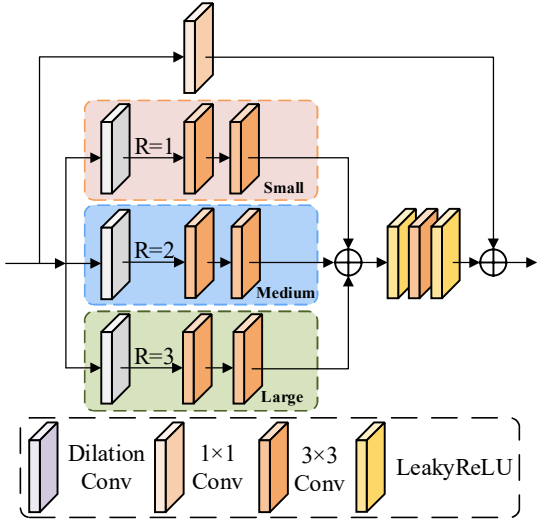

**Fig. 2.** The multi-scale residual module of the proposed MAE-GAN. (Large: Large receptive field feature information flow. Medium: Medium receptive field feature information flow. Small: Small receptive field feature information flow)

$$DAM(x) = S[C(x)] \quad (1)$$

where *DAM* represents the dual attention module, $C(x)$ represents the channel attention module, $S(y)$ represents the spatial attention module, $x$ represents the input to the channel attention module, and $y$ represents the input to the spatial attention module, i.e. $y = C(x)$. The channel attention module $C(x)$ can be expressed as:

$$C(x) = x \otimes \delta\left\{Conv_{1\times1}[GAP(x)]\right\} \quad (2)$$

where $\delta$ represents the sigmoid activation function, while $Conv_{1\times1}$ and *GAP* represent a 1×1 convolutional layer and global average pooling layer, respectively. The channel attention module $S(y)$ can be represented as:

$$S(y) = x \otimes \delta\left\{Conv_{3\times3}\left\{\left[Conv_{1\times1}\begin{bmatrix}CMP(y)\\ \circ CAP(y)\\ \circ Up(Conv_{3\times3\times3}(Down(y)))\end{bmatrix}\right]\right\}\right\} \quad (3)$$

where *CMP* represents a global maximum pooling layer in channel dimensions, *CAP* represents a global average pooling layer in channel dimensions, *Down* represents downsampling operation, *Up* represents upsampling operation, and $Conv_{3\times3\times3}$ represents three cascaded 3×3 convolutional layers.

The complete structure of the MAE-GAN's generative network is shown in Fig. 1. The MAE-GAN's generative network consists of an encoder and a decoder. The encoder is composed of a 3×3 convolutional layer with LeakyReLU activation, two proposed multi-scale residual modules (as shown in Fig. 2), two proposed dual attention modules (as shown in Fig. 3), two ResNet-D modules, and three cascaded dilated convolution layers. Low-resolution post-stack seismic profiles are first processed by a 3×3 convolutional layer with LeakyReLU activation to extract local feature information, increasing the number of feature map channels to 32. The multi-scale residual modules are used to extract multi-scale information of the feature maps, doubling the channel number of the input feature map. To minimize the interference of noise, our designed dual attention modules assign greater weight to the extracted multi-scale critical information. To better extract low-frequency information of seismic profiles, we use ResNet-D instead of pooling layers for downsampling. The structure of ResNet-D, as shown in Fig. 1, consists of two branches: the first branch is composed of two 1×1 convolutional layers and one 3×3 convolutional layer with the stride of 2, and the second branch consists of an average pooling layer followed by a 1×1 convolutional layer. ResNet-D enhances feature information retention and transmission by introducing an average pooling layer in the traditional ResNet downsampling phase. Unlike traditional methods that use pooling or convolution strides for downsampling, ResNet-D smoothly reduces the resolution of the feature map by average pooling before applying convolution, thereby reducing information loss and enhancing the network's ability to extract detailed information from the feature maps. After downsampling, the size of the output feature map is reduced to half the size of the input feature map. The process continues with the same multi-scale residual modules, dual attention modules, and ResNet-D, further extracting features. The size of the output feature map is reduced to one-quarter of the input low-resolution post-stack profile size, and the number of channels increases to 128. To further increase the receptive field and extract more accurate low-frequency information, we designed a dilated convolution group, as shown in Fig. 1, composed of three cascaded 3×3 dilated convolutions. The output feature map of the dilated convolution group serves as the input feature map for the decoder.

The designed generative network's decoder consists of two upsampling layers, two 1×1 convolutional layers, three multi-scale residual modules, three dual attention modules, a pixel shuffle layer, and one 3×3 convolutional layer with a tanh activation function. To better reconstruct high-resolution seismic profiles, we use bilinear interpolation for upsampling, increasing the feature map size to half the size of the input low-resolution seismic profile. Dual attention modules and multi-scale residual modules are then used to assign better weights to important features and extract multi-scale high-frequency information. To better fuse the low-frequency information extracted by the encoder with the high-frequency information extracted by the decoder, we use skip connections to fuse the output feature maps of the decoder's multi-scale residual modules with the output feature maps of the encoder's ResNet. Subsequently, 1×1 convolution layers are used to adjust the number of channels. Similarly, the same upsampling layers, dual attention modules, multi-scale residual modules, skip connections, and 1×1 convolution layers are applied at the next scale. Afterward, dual attention modules and multi-scale residual modules are again used to further extract deep feature information. Finally, a pixelshuffle layer increases the feature map size to twice that of the input low-resolution post-stack seismic profile, and a 3×3 convolutional layer with Tanh function is used to obtain a high-resolution seismic profile.

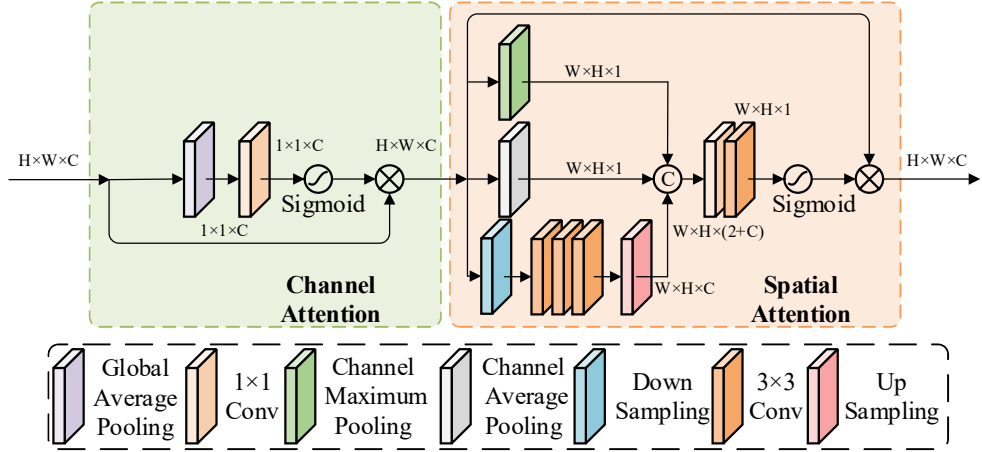

**Fig. 3.** The dual attention module of the proposed MAE-GAN.

### B. The architecture of Adversarial Network in MAE-GAN

The proposed MAE-GAN's adversarial network, as shown in Fig. 4, is based on a Markovian discriminator that enhances discrimination performance. Furthermore, the adversarial network employs fully convolutional layers instead of fully connected layers, significantly reducing computation. The adversarial network consists of three 3×3 convolutions with a stride of 2, each featuring LeakyReLU activation, and one independent 3×3 convolution also with a stride of 2. The output feature map size of the three 3×3 convolutions is reduced to one-eighth the size of the input post-stack seismic profile, while the output feature map size of the final independent 3×3 convolution is reduced to one-sixteenth the size of the input profile. The number of channels in the output feature map is adjusted to 1. The output single-channel feature map of the last independent 3×3 convolution serves as the discrimination score matrix that determines whether the input image to the adversarial network is a high-resolution post-stack seismic profile from the dataset or a high-resolution post-stacked seismic profile reconstructed by the generative network.

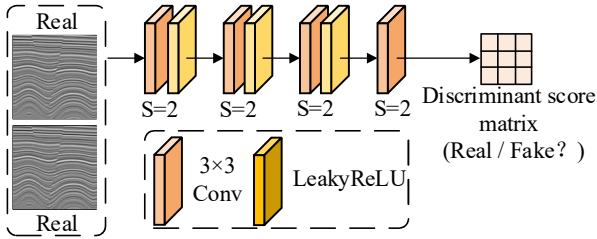

**Fig. 4.** The adversarial network structure of the proposed MAE-GAN.

### C. The Loss Function of MAE-GAN

To further enhance the performance of post-stack seismic profiles super-resolution reconstruction, we propose an improved loss function by incorporating an edge information preservation loss function into the conventional loss function of GAN. The loss function of the proposed MAE-GAN can be expressed as:

$$Loss = 0.01 L_{adv} + 2 \times 10^{-8} L_{tv} + L_{mse} + L_{edge} \quad (4)$$

where $L_{adv}$ represents the adversarial loss function, $L_{tv}$ represents the total variation loss function, $L_{mse}$ represents the mean squared error loss function, and $L_{edge}$ represents the edge information preservation loss function. The adversarial loss function $L_{adv}$ can be expressed as:

$$L_{adv} = E[\log D(y)] + E\{\log[1 - D(G(x))]\} \quad (5)$$

where $x$ is the input low-resolution stacked seismic section, $y$ is the high-resolution stacked seismic section in the training set, $G$ is the generator network, $D$ is the adversarial network, $G(x)$ is the reconstructed high-resolution post-stack seismic section, represents the mean operation.

The total variation loss function $L_{tv}$ can reduce noises by smoothing operations, it can be expressed as:

$$L_{tv} = \sum_{i=1}^{W} \sum_{j=1}^{H} \left( \begin{array}{c} (G(x)_{i,j+1} - G(x)_{i,j})^2 \\ + (G(x)_{i+1,j} - G(x)_{i,j})^2 \end{array} \right)^{\frac{\beta}{2}} \quad (6)$$

where $\beta$ is set to 2.

The mean squared error loss function $L_{mse}$ is used to measure the detailed differences between the reconstructed post-stack seismic profiles and the high-quality post-stack seismic profiles in the test set, it can be expressed as:

$$L_{mse} = \sum_{i=1}^{W} \sum_{j=1}^{H} \left\| G_{i,j}(x) - y_{i,j} \right\|_2 \quad (7)$$

The edge information preservation loss function $L_{edge}$ is used to measure the differences in edge textures between the reconstructed high-quality post-stack seismic profiles and the high-quality post-stack seismic profiles in the test set. The edge information preservation loss function $L_{edge}$ can be expressed as:

$$L_{edge} = \sum_{i=1}^{W} \sum_{j=1}^{H} \left\| Sobel[G_{i,j}(x)] - Sobel[y_{i,j}] \right\|_1 \quad (8)$$

where $Sobel(\ )$ represents the edge detection operation [40].

III. TRAINING DATASET AND EXPERIMENTAL ENVIRONMENT

*A. Construction of Dataset*

The process for creating paired low-resolution and high-resolution post-stack seismic profiles is shown in Fig. 5 and can be divided into the following steps:

*Step 1:* Create a one-dimensional synthetic reflectivity model

Randomly generate reflectivity values within a specific depth range. These values are distributed within the range of [-1, 1], used to simulate the reflective characteristics inside geological layers.

*Step 2:* Apply Gaussian deformation to create folded structures

Use a two-dimensional Gaussian function to deform the reflectivity model, simulating the bending and folding of geological layers. This step creates complex shapes of geological structures by altering the vertical and horizontal positions of the layers.

*Step 3:* Add planar shearing to simulate folds and faults

Apply planar shearing to the deformed model to further increase its complexity. This process simulates the more extensive displacement and deformation of layers under geological pressures, such as folds and faults.

*Step 4:* Convolve with the high-frequency Ricker wavelet

First, convolve the reflectivity model with the high-frequency Ricker wavelet to generate three-dimensional high-frequency seismic data. This step simulates the propagation of high-frequency seismic waves in the subsurface medium, reflecting detailed structural features of the layers. Then, extract the two-dimensional high-frequency seismic profile from the three-dimensional high-frequency seismic data, which serves as the ground truth for further analysis.

*Step 5:* Convolve with the low-frequency Ricker wavelet and add noise

First, convolve the reflectivity model with the low-frequency Ricker wavelet to generate three-dimensional low-frequency seismic data. Low-frequency seismic waves can better penetrate deep geological structures but capture less detailed geological information. To enhance the model's realism, further add random colored noise to the three-dimensional low-frequency seismic data. This step simulates the inevitable environmental and instrumental noise in the field seismic data collection process. Finally, extract the two-dimensional noisy low-frequency seismic profile from the noise-added three-dimensional low-frequency seismic data, serving as the model input.

*Step 6:* Downsampling operation

To simulate low-resolution seismic data, downsample the two-dimensional noisy low-frequency seismic profile. This step reduces the dimensions and resolution of the data by extracting pixels in both horizontal and vertical directions.

The downsampled two-dimensional noisy geological profile serves as the low-resolution post-stack seismic profile, and the extracted two-dimensional seismic profile from the three-dimensional high-frequency seismic data serves as the high-resolution post-stack seismic profile.

Using this data construction method, we generated 1600 pairs of low-resolution/high-resolution seismic profiles. Of these, 70% (1120 pairs) of the seismic profiles were used for training, 10% (160 pairs) were used for validation, and 20% (320 pairs) were used for testing.

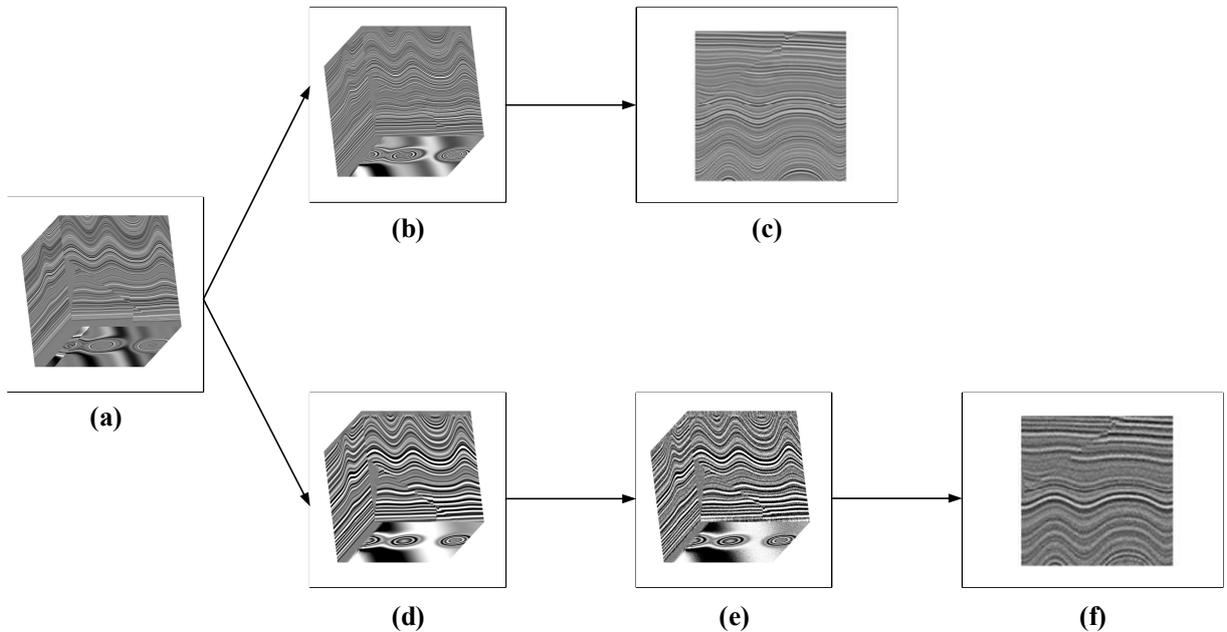

**Fig. 5.** Process of creating synthetic training datasets. (a) Reflectivity model. (b) High-resolution 3-dimensional (3D) seismic data. (c) High-resolution seismic profile. (d) Noise-free Low-resolution 3D seismic data. (e) Noisy low-resolution 3D seismic data. (f) Low-resolution seismic profile.

## B. Training Details and Experimental Environment

During the training phase, we normalize the post-stack profile data to the range of [0, 1] so that the network can quickly learn the relationship between low-resolution and high-resolution seismic data. The normalization operation can be expressed as:

$$x_n = \frac{x - x_{min}}{x_{max} - x_{min}} \quad (9)$$

where $x_n$ represents the normalized post-stack profile, and $x_{max}$ and $x_{min}$ represent the maximum and minimum values of each seismic profile, respectively.

To ensure fairness and consistency in our experiment, we uniformly applied the Adam optimizer to optimize the loss function, with parameters $\beta_1$ = 0.5 and $\beta_2$ = 0.999 used to adjust the decay rates of the exponential moving averages. Our experiment involved training for 200 epochs with a learning rate of $1 \times 10^{-4}$. The experiments were conducted on Ubuntu 20.04 using the PyTorch deep learning framework, with an NVIDIA GeForce RTX 3090 GPU.

To evaluate the denoising performance and super-resolution reconstruction capabilities, we adopted the peak signal-to-noise ratio (PSNR) [41] and structural similarity (SSIM) [42] as evaluation metrics. The PSNR can be stated as follows:

$$PSNR = 10 \times log_{10}(\frac{MAX^2}{MSE}) \quad (10)$$

where *MAX* represents the maximum amplitude of the post-stack profile, and *MSE* represents the mean square error. MSE can be written as:

$$MSE = \frac{1}{mn} \sum_{i=0}^{m-1} \sum_{j=0}^{m-1} \|I_R(i,j) - I_G(i,j)\|^2 \quad (11)$$

where *m* and *n* represent the dimensions of the post-stack seismic profile. $I_R$ represents the reconstructed high-resolution seismic profile, while $I_G$ represents the original high-resolution seismic profile. A higher PSNR value indicates that the reconstructed seismic profile is more similar to the original high-resolution seismic profile, with less noise. SSIM can be represented as follows:

$$SSIM(x,y) = \frac{(2\mu_x\mu_y + c_1)(2\sigma_{xy} + c_2)}{(\mu_x^2 + \mu_y^2 + c_1)(\sigma_x^2 + \sigma_y^2 + c_2)} \quad (12)$$

where $\mu_x$, $\mu_y$ represents the mean of the two post-stack profiles, $\sigma_x^2$, $\sigma_x^2$ and $\sigma_y^2$, $\sigma_y^2$ represent the variance of the two post-stack profiles, while $\sigma_{xy}$ represents the covariance between the reconstructed seismic profile and the original high-resolution seismic profile. The constants $c_1$ and $c_2$ ($c_1=(k_1L)$ and $c_2=(k_2L)$) are used to prevent the denominator from becoming zero, which would cause the model to be unstable. $L=2^{bit}-1$ represents the amplitude dynamic range. Default values for $k_1$ and $k_2$ are set at 0.01 and 0.03, respectively. As the SSIM value increases, the reconstructed seismic profile becomes closer to the original high-resolution seismic profile.

## IV. Synthetic Example

### A. Competitive Methods

In this paper, we compare the proposed MAE-GAN with two other advanced methods: U-Net and SRGAN. Both U-Net and SRGAN are classic convolutional neural network methods for processing images. To ensure fairness, we set all training hyperparameters of these two methods to be the same as those of MAE-GAN. Additionally, we use the dataset constructed in part three to train, validate, and test U-Net, SRGAN, and our proposed MAE-GAN.

### B. Comparison of Super-Resolution Reconstruction Performance

We randomly selected a low-resolution seismic profile from the test set to test the super-resolution reconstruction performance of U-Net, SRGAN, and the proposed MAE-GAN. Fig. 6 shows the test results, where Fig. 6(a) and (e) show the input noisy low-resolution seismic profile and the paired noise-free high-resolution seismic profile, respectively. Fig. 6(b), (c), and (d) respectively present the high-resolution seismic profiles reconstructed by U-Net, SRGAN, and the proposed MAE-GAN. As shown in Fig. 6, the U-Net, SRGAN and the proposed MAE-GAN methods effectively enhance the resolution of the post-stack seismic profiles. However, the high-resolution seismic profile reconstructed by the SRGAN method contains noise (as indicated by the red and yellow arrows). The weak signals in the high-resolution seismic profile reconstructed by the U-Net method are not effectively restored (as indicated by the yellow arrow). Compared to the other methods, the proposed MAE-GAN method better improves the resolution of the post-stack seismic profiles and effectively removes the noise in the seismic profiles. The high-resolution post-stack seismic profile reconstructed by the proposed MAE-GAN method exhibits clearer geological structures, such as faults.

To better compare the super-resolution reconstruction performance of different methods, we conducted a single-trace amplitude analysis on a post-stack seismic profile reconstructed by the U-Net, SRGAN, and the proposed MAE-GAN methods. The test subject chosen is the low-resolution post-stack profile randomly extracted as mentioned earlier. The result of the single-trace amplitude comparison is shown in Fig. 7. The four curves in Fig. 7 represent the noise-free high-resolution post-stack seismic profile from the test set, the seismic profile reconstructed by the U-Net method, the seismic profile reconstructed by the SRGAN method, and the seismic profile reconstructed by the proposed MAE-GAN method. As illustrated in Fig. 7, compared to the U-Net and SRGAN methods, the single-trace amplitude curve of the proposed MAE-GAN method is closer to that of the noise-free high-resolution post-stack seismic profile. This demonstrates that the MAE-GAN method possesses superior super-resolution reconstruction performance.

Additionally, to further validate the super-resolution performance of different methods, we also compared the frequency spectra of the seismic profiles reconstructed by these methods. As shown in the results in Fig. 8, the U-Net method, SRGAN method, and our proposed MAE-GAN method all effectively increase the main frequency of the seismic data and broaden the bandwidth. However, compared to the frequency spectra of the U-Net and SRGAN methods, the power spectrum of our proposed MAE-GAN method is closer to that of a noise-free high-resolution seismic profile. Therefore, the MAE-GAN method exhibits superior super-resolution reconstruction performance.

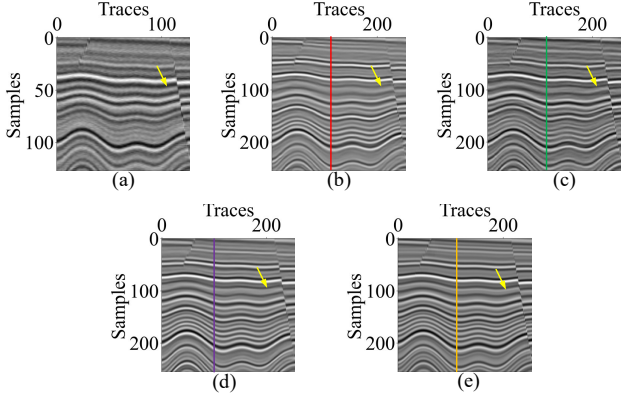

**Fig. 6.** Test results on synthetic post-stack seismic profiles: (a) noisy low-resolution seismic profile, (b) recovered seismic image by U-Net, (c) recovered seismic image by SRGAN, (d) recovered seismic image by proposed MAE-GAN, (e) the paired noise-free high-resolution seismic profile.

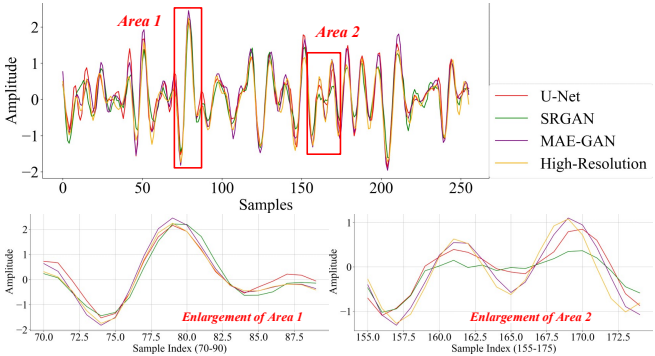

**Fig. 7.** Single-trace amplitude comparison curve. Pink curve: seismic profile reconstructed by the U-Net method [Fig. 6(b)]. Green curve: seismic profile reconstructed by the SRGAN method [Fig. 6(c)]. Blue curve: seismic profile reconstructed by the proposed MAE-GAN method [Fig. 6(d)]. Orange curve: noise-free high-resolution post-stack seismic profile [Fig. 6(e)].

To better validate the superiority of the proposed MAE-GAN method, we conducted quantitative experiments in the entire test set. The experimental results are shown in Table I. The PSNR values for the U-Net method, SRGAN method, and the proposed MAE-GAN method are 22.383 dB, 21.682 dB, and 25.663 dB, respectively. The SSIM values for the U-Net method, SRGAN method, and the proposed MAE-GAN method are 0.792, 0.794, and 0.854, respectively. Our proposed MAE-GAN method has the largest PSNR and SSIM values, which indicates that MAE-GAN can effectively enhance the resolution of post-stack profiles, remove noises, and reconstruct seismic profiles that are closer to the noise-free high-resolution post-stack seismic profiles.

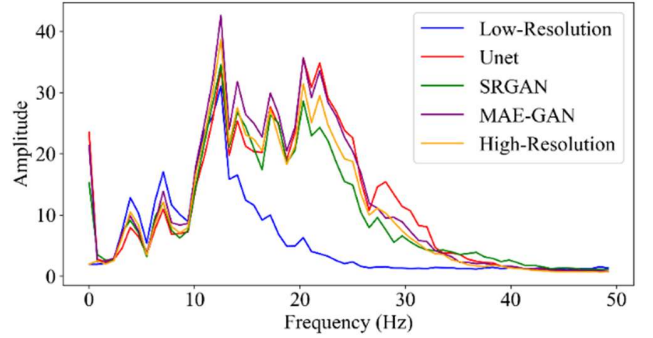

**Fig. 8.** Spectrum analysis of Synthetic seismic field.

TABLE I
COMPARISON OF SUPER-RESOLUTION RECONSTRUCTION
PERFORMANCE OF DIFFERENT METHODS

|      | U-Net  | SRGAN  | MAE-GAN |
|------|--------|--------|---------|
| PSNR | 22.383 | 21.682 | 25.663  |
| SSIM | 0.792  | 0.794  | 0.854   |

## V. FIELD EXAMPLE

In this section, we used field seismic data from the Netherlands F3 block in the North Sea oil field as a test subject to validate the super-resolution reconstruction performance of different methods. We input the field seismic data into models that were trained with simulated data for the experiments. The experimental results are shown in Fig. 9. Fig. 9(a) and (e) display the input low-resolution field data. The second, third, and fourth columns of Fig. 9 respectively represent the high-resolution seismic data reconstructed by the U-Net method, SRGAN method, and the proposed MAE-GAN method. Below each reconstructed high-resolution seismic profile is a close-up of the corresponding area highlighted in red. As depicted in Fig. 9, the U-Net, SRGAN, and proposed MAE-GAN methods effectively restored the stratigraphic structures and eliminated noises. However, the data reconstructed by the U-Net and SRGAN methods did not preserve weaker amplitude signals well (as indicated by the yellow arrows). In comparison to the U-Net and SRGAN methods, MAE-GAN method restored high-resolution data from different field seismic data with clearer stratigraphic structures and improved recovery of weak amplitude signals. Therefore, the proposed MAE-GAN method demonstrates the best super-resolution reconstruction performance.

Additionally, we compared the single trace amplitude curves of the input field seismic data with the reconstructed seismic data. Fig. 10 displays the single trace amplitude curves extracted from the input field seismic data and the high-resolution seismic data reconstructed by the proposed MAE-GAN method. Fig. 10(a) shows a randomly selected single trace from Figures 9(a) and (d), while Fig. 10(b) is from Fig. 9(e) and (h). The blue curve represents the single trace of the field seismic data, and the red curve represents the single trace of the high-resolution seismic data reconstructed by the proposed MA-GAN method. From Fig. 8, it can be seen that the single trace curves of the high-resolution seismic data reconstructed by the proposed MAE-GAN method have the same waveform characteristics as the input field seismic data traces and the MAE-GAN results exhibit higher frequencies, indicating that MAE-GAN preserved more detailed features.

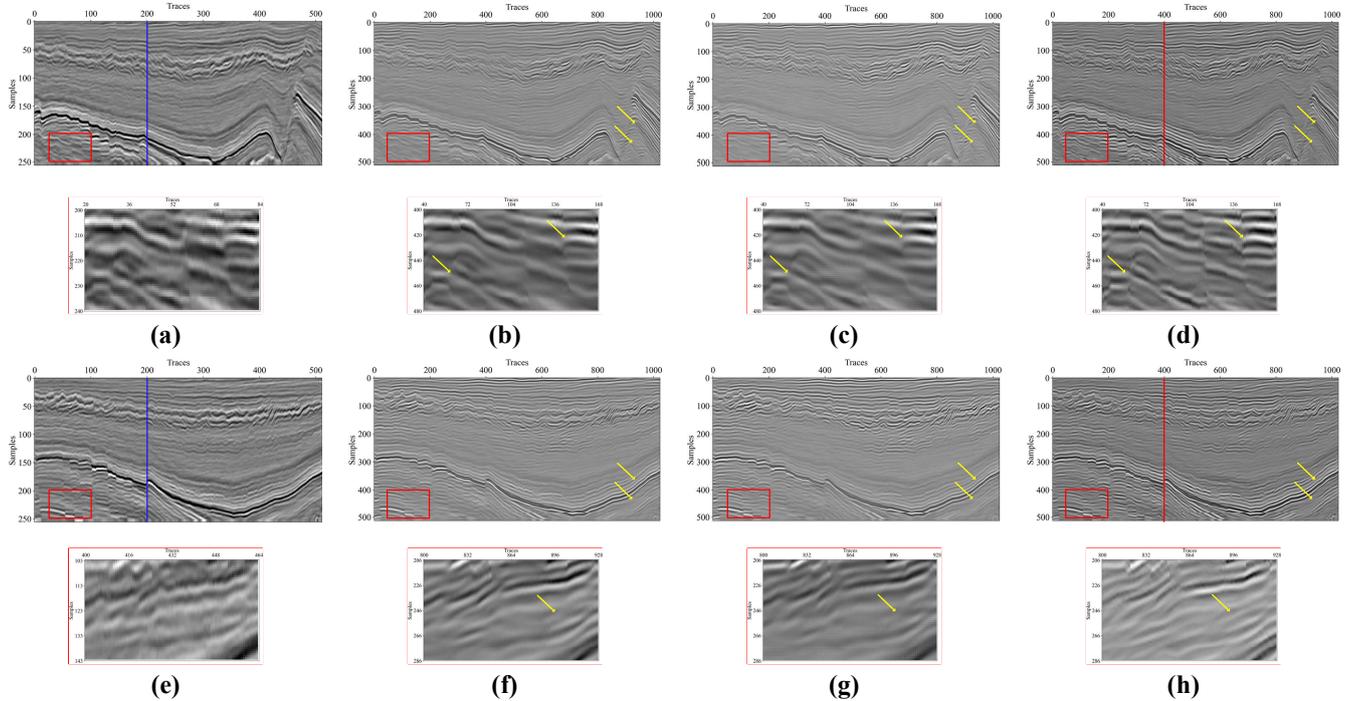

**Fig. 9.** Comparison of super-resolution reconstruction results of different methods in the Netherlands F3 block in the North Sea oil field. (a) and (e) low-resolution field seismic datas, (b) and (f) seismic profiles reconstructed by U-Net method, (c) and (g) seismic profiles reconstructed by SRGAN method, (d) and (h) seismic profile reconstructed by the proposed MAE-GAN method.

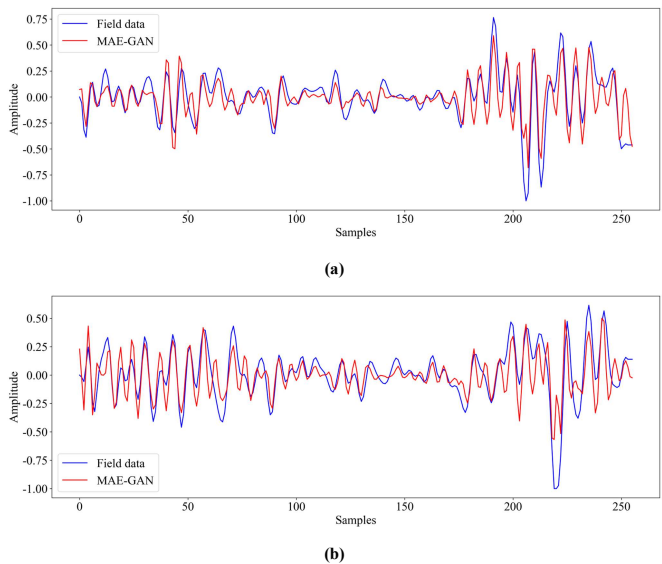

**Fig. 10.** Comparison of single-trace amplitude curves of seismic data after reconstruction by different methods. Blue curve: Input field seismic data [Fig. 9(a) and (e)]. Red curve: seismic profile reconstructed by the proposed MAE-GAN method [Fig. 9(d) and (h)].

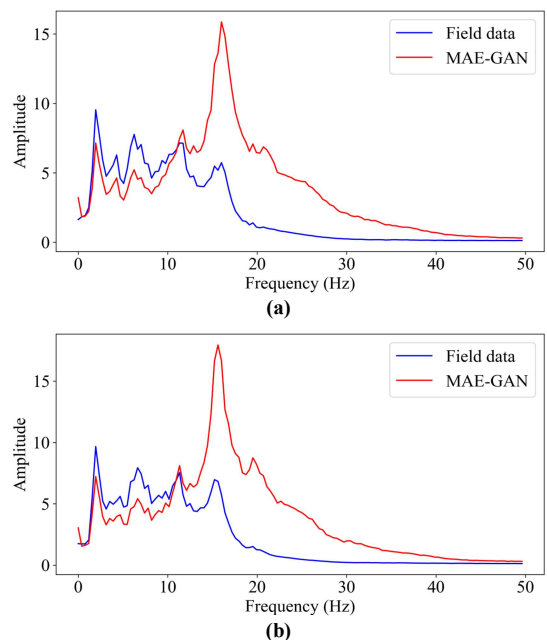

**Fig. 11.** Spectrum analysis of seismic data reconstructed by different methods in three field seismic images: (a) first field

data, (b) second field data. Blue curve: Input field seismic data [Fig. 9(a) and (e)]. Red curve: seismic profile reconstructed by the proposed MAE-GAN method [Fig. 9(d) and (h)].

Finally, we also conducted a spectral analysis, the results of which are shown in Fig. 11. Fig. 11(a) shows the spectral curves of Fig. 9(a) and (d), while Fig. 11(b) shows the spectral curves of Fig. 9(e) and (h). In the spectral graphs, the amplitude of each frequency is obtained by averaging all traces in the seismic profile. The blue and red curves respectively represent the spectral curves of the input field seismic data and the high-resolution seismic profiles reconstructed by the proposed MAE-GAN method. The results show that the frequency band of the high-resolution seismic profiles reconstructed by the proposed MAE-GAN method is wider than that of the input field seismic profiles, especially in the high-frequency range.

## VI. DISCUSSION

### A. Generalization

In the field of seismic data processing, super-resolution reconstruction techniques have demonstrated their ability to enhance the resolution of seismic profiles and recover subtle geological structures. In particular, methods based on GANs have shown superior performance in improving the resolution and signal-to-noise ratio of post-stack seismic profiles. However, the success of these methods often relies on the quality and diversity of the training datasets, and their generalization ability to handle new data significantly different from the training set has not been fully verified. In light of this, this section uses a new seismic data as the subject to assess the effectiveness and robustness of the proposed MAE-GAN model in handling diverse and unseen seismic datasets, thus exploring the model's generalization performance. We continue to use the model trained with simulated low-resolution to high-resolution seismic profiles to process a seismic profile with completely different stratigraphic features than previous data. The experimental results are shown in Fig. 12. Fig. 12(a) shows the Parihaka seismic data, provided by the New Zealand Crown Minerals, while Fig. 12(b), (c), and (d) are the seismic data reconstructed by the U-Net, SRGAN, and MAE-GAN methods, respectively.

In Fig. 12(b) (the reconstructed result of the U-Net method) and Fig. 12(c) (the reconstructed result of the SRGAN method), it can be observed that the details of the geological structures are blurry, with no clear stratification lines, indicating that these models may underperform in terms of high-frequency details. Furthermore, in the U-Net reconstruction, layer discontinuities or inconsistencies can be seen, lacking stratigraphic continuity. This is due to the U-Net network relying solely on convolutional layers to extract features, resulting in the network's inability to effectively learn the complex mapping relationships between seismic data. In the SRGAN method reconstruction, obvious artifacts and noise can be seen (indicated by yellow arrows), which is due to the limitations of the receptive field in the SRGAN model, leading to the network learning incorrect geological features. The proposed MAE-GAN method effectively enhances the resolution of the seismic data and eliminates noise. Compared to U-Net and SRGAN methods, the high-resolution seismic data reconstructed by the proposed MAE-GAN method retains more details, appears more natural visually, and has no noises and artifacts.

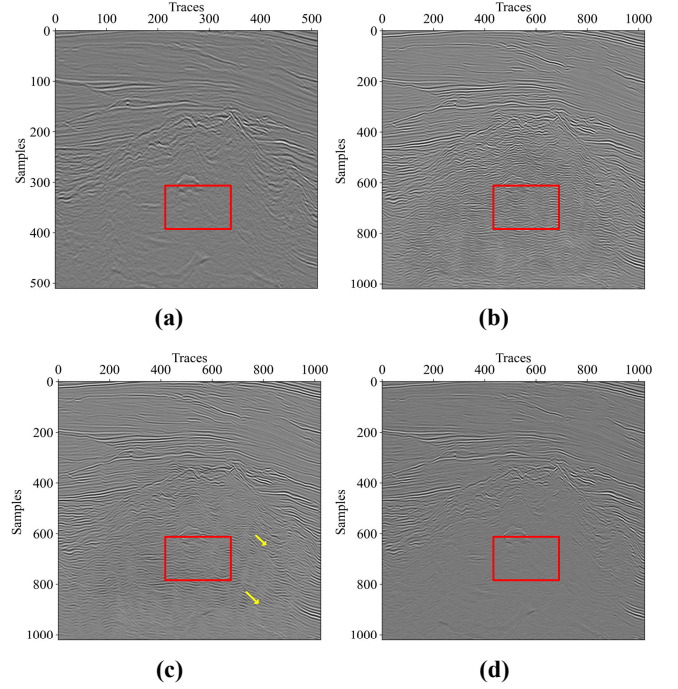

**Fig. 12.** Comparison of super-resolution reconstruction results of different methods in New Zealand seismic data. (a) low-resolution field seismic data, (b) seismic profiles reconstructed by the U-Net method, (c) seismic profile reconstructed by the SRGAN method, (d) seismic profile reconstructed by the proposed MAE-GAN method.

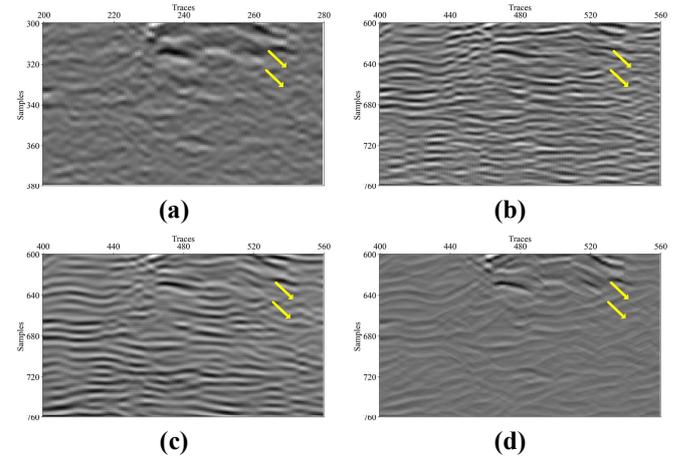

**Fig. 13.** Enlarged profile of the red area in Fig. 12. (a) low-resolution field seismic data, (b) seismic profiles reconstructed by the U-Net method, (c) seismic profile reconstructed by the SRGAN method, (d) seismic profile reconstructed by the proposed MAE-GAN method.

To further compare the signal recovery capabilities of different methods, we also selected one area for magnification. As shown in Fig. 13, (a), (b), (c), and (d) are the magnified profiles of the red area in Fig. 12(a), (b), (c), and (d), respectively. The results show that the U-Net and SRGAN methods effectively enhance the resolution and signal-to-noise ratio of the data, but they both introduce artifacts, resulting in false coherent axes in the reconstructed seismic data. The proposed MAE-GAN method effectively avoids these issues and reconstructs high-resolution data that is clearer, more realistic, and of higher perceptual quality. Therefore, the proposed MAE-GAN method has the best generalization performance.

*B. Ablation experiment*

To better demonstrate the effectiveness of each module in the proposed MAE-GAN method, we designed ablation experiments. We continue to use simulated low-resolution to high-resolution seismic data as the experimental subjects, and conducted the following four experiments:

- w/o MRM: Without the multi-scale residual module
- w/o DAM: Without the dual attention module
- w/o EDGE: Without the edge information preservation loss function
- MAE-GAN: The complete network based on the proposed modules

The results in Table II show that the PSNR values for w/o MRM, w/o DAM, w/o EDGE, and MAE-GAN are 22.383, 0.9124, 0.9225, and 0.9126, respectively. The SSIM values for w/o MRM, w/o DAM, w/o EDGE, and MAE-GAN are 0.8263, 0.9124, 0.9225, and 0.9126, respectively. Considering all modules, the proposed MAE-GAN network demonstrated the highest SSIM and PSNR values. This highlights the significant contributions of each module in the MAE-GAN network to the post-stack seismic resolution reconstruction process.

TABLE II
EVALUATION RESULTS OF EACH MODULE

|      | w/o MRM | w/o DAM | w/o EDGE | MAE-GAN |
|------|---------|---------|----------|---------|
| PSNR | 21.554  | 22.425  | 24.840   | 25.663  |
| SSIM | 0.794   | 0.834   | 0.841    | 0.854   |

VII. CONCLUSION

In this paper, we proposed a multi-scale attention encoder-decoder network based on generative adversarial network, MAEGAN, for simultaneous super-resolution reconstruction and denoising of post-stack seismic profiles. In the generative of MAEGAN, multi-scale residual modules are employed to extract multi-scale stratigraphic feature information. Dual attention modules enhance the capability to extract important feature information by increasing the weight of important features. The adversarial network based on Markov discriminators effectively improves the network's discriminative performance, thereby assisting in enhancing the network's super-resolution reconstruction and denoising capabilities. The introduction of the edge information preservation loss function better retains features such as faults, effectively improving the perceived quality of the reconstructed seismic profiles. We conducted comparative experiments on synthetic and field data, and the results show that the performance of the proposed MAE-GAN method surpasses that of two existing super-resolution methods. Additionally, the generalizability of the MAE-GAN method and the effectiveness of its components were validated in the discussion section.

REFERENCES

[1] J. Chen et al., "Efficient Seismic Data Denoising via Deep Learning With Improved MCA-SCUNet," *IEEE Trans. Geosci. Remote Sens.*, vol. 62, pp. 1-14, 2024, doi: 10.1109/TGRS.2024.3355972.
[2] H. Wang, J. Lin, X. Dong, S. Lu, Y. Li, and B. Yang, "Seismic velocity inversion transformer," *Geophysics*, vol. 88, no. 4, pp. R513–R533, Jul. 2023, doi: https://doi.org/10.1190/geo2022-0283.1.
[3] C. Liu et al., "Deep learning-based point-spread function deconvolution for migration image deblurring," *Geophysics*, vol. 87, no. 4, pp. S249–S265, Jun. 2022, doi: https://doi.org/10.1190/geo2020-0904.1.
[4] S. D. Billings, B. R. Minty, and G. N. Newsam, "Deconvolution and spatial resolution of airborne gamma-ray surveys," *Geophysics*, vol. 68, no. 4, pp. 1257–1266, Jul. 2003, doi: https://doi.org/10.1190/1.1598118.
[5] G. Tian, Y. Zhao, W. Zhang, and C. Yang, "The inverse Q filtering method based on a novel variable stability factor," *Geophysics*, vol. 88, no. 3, pp. V207–V214, Apr. 2023, doi: https://doi.org/10.1190/geo2022-0088.1.
[6] Y. Zhao, "An inverse Q filtering method with adjustable amplitude compensation operator," *J Appl Geophy.*, vol. 215, pp. 105111–105111, Aug. 2023, doi: https://doi.org/10.1016/j.jappgeo.2023.105111.
[7] J. Yang, J. Huang, H. Zhu, G. McMechan, and Z. Li, "An Efficient and Stable High-Resolution Seismic Imaging Method: Point-Spread Function Deconvolution," *Journal of geophysical research. Solid earth*, vol. 127, no. 7, Jul. 2022, doi: https://doi.org/10.1029/2021jb023281.
[8] G. Xin, Gao Jian-Hu, Yin Xun-De, Y. Xue-Shan, Wang Hong-Qiu, and L. Sheng-Jun, "Seismic high-resolution processing method based on spectral simulation and total variation regularization constraints," *Appl. Geophys.*, vol. 19, no. 1, pp. 81–90, Mar. 2022, doi: https://doi.org/10.1007/s11770-022-0927-5.
[9] Z. Chen, Y. Wang, X. Chen, and J. Li, "High-resolution seismic processing by Gabor deconvolution," *J. Geophys. Eng.*, vol. 10, no. 6, Oct. 2013, doi: https://doi.org/10.1088/1742-2132/10/6/065002.
[10] Y. Wang, G. Zhang, H. Li, W. Yang, and W. Wang, "The high-resolution seismic deconvolution method based on joint sparse representation using logging–seismic data," *Geophys. Prospect.*, vol. 70, no. 8, pp. 1313–1326, Jul. 2022, doi: https://doi.org/10.1111/1365-2478.13232.
[11] G. Zhang, X. Wang, and Z. He, "A stable and self-adaptive approach for inverse Q-filter," *Appl. Geophys.*, vol. 116, pp. 236–246, May 2015, doi: https://doi.org/10.1016/j.jappgeo.2015.03.012.
[12] Y. Wang, "Inverse Q-filter for seismic resolution enhancement," *Geophysics*, vol. 71, no. 3, pp. V51–V60, May 2006, doi: https://doi.org/10.1190/1.2192912.
[13] S. Braga and F. Moraes, "High-resolution gathers by inverse Q filtering in the wavelet domain," *Geophysics*, vol. 78, no. 2, pp. V53–V61, Mar. 2013, doi: https://doi.org/10.1190/geo2011-0508.1.
[14] Niloofar Alaei, Amin Roshandel Kahoo, Abolghasem Kamkar Rouhani, and M. Soleimani, "Seismic resolution enhancement using scale transform in the time-frequency domain," *Geophysics*, vol. 83, no. 6, pp. V305–V314, Nov. 2018, doi: https://doi.org/10.1190/geo2017-0248.1.
[15] O. Ovcharenko, V. Kazei, M. Kalita, D. Peter, and T. Alkhalifah, "Deep learning for low-frequency extrapolation from multioffset seismic data," *Geophysics*, vol. 84, no. 6, pp. R989–R1001, Nov. 2019, doi: https://doi.org/10.1190/geo2018-0884.1.
[16] H. Kaur, N. Pham, and S. Fomel, "Seismic data interpolation using deep learning with Generative Adversarial Networks," *Geophys. Prospect.*, Nov. 2020, doi: https://doi.org/10.1111/1365-2478.13055.
[17] J. Li, X. Wu and Z. Hu, "Deep Learning for Simultaneous Seismic Image Super-Resolution and Denoising," *IEEE Trans. Geosci. Remote Sens.*, vol. 60, pp. 1-11, 2022, doi: 10.1109/TGRS.2021.3057857.

[18] Y. Gao, D. Zhao, T. Li, G. Li and S. Guo, "Deep Learning Vertical Resolution Enhancement Considering Features of Seismic Data," *IEEE Trans. Geosci. Remote Sens.*, vol. 61, pp. 1-13, 2023, Art no. 5900913, doi: 10.1109/TGRS.2023.3234617.

[19] Y. Wang, J. Xu, Z. Zhao, Y. Gao, and H. Zhang, "Structurally-Constrained Unsupervised Deep Learning for Seismic High-Resolution Reconstruction," *IEEE Trans. Geosci. Remote Sens.*, vol. 62, pp. 1–15, Jan. 2024, doi: https://doi.org/10.1109/tgrs.2023.3340888.

[20] Adnan Hamida, Motaz Alfarraj, A. A. Al-Shuhail, and S. A. Zummo, "Facies-Guided Seismic Image Super-Resolution," *IEEE Trans. Geosci. Remote Sens.*, vol. 61, pp. 1–13, Jan. 2023, doi: https://doi.org/10.1109/tgrs.2023.3289151.

[21] D. Zeng, Q. Xu, S. Pan, G. Song, and F. Min, "Seismic image super-resolution reconstruction through deep feature mining network," *Appl. Intell.*, vol. 53, no. 19, pp. 21875–21890, Jun. 2023, doi: https://doi.org/10.1007/s10489-023-04660-y.

[22] H. Zhang, Tariq Alkhalifah, Y. Liu, C. Birnie, and X. Di, "Improving the Generalization of Deep Neural Networks in Seismic Resolution Enhancement," *IEEE Trans. Geosci. Remote Sens.*, vol. 20, pp. 1–5, Jan. 2023, doi: https://doi.org/10.1109/lgrs.2022.3229167.

[23] Y. Choi, Y. Jo, Soon Jee Seol, J. Byun, and Y. Kim, "Deep learning spectral enhancement considering features of seismic field data," *Geophysics*, vol. 86, no. 5, pp. V389–V408, Aug. 2021, doi: https://doi.org/10.1190/geo2020-0017.1.

[24] R. Zhou, C. Zhou, Y. Wang, X. Yao, G. Hu, and F. Yu, "Deep Learning With Fault Prior for 3-D Seismic Data Super-Resolution," *IEEE Trans. Geosci. Remote Sens.*, vol. 61, pp. 1–16, Jan. 2023, doi: https://doi.org/10.1109/tgrs.2023.3262884.

[25] L. Trevisan, B. Augusto, Harlen Costa Batagelo, and João Paulo Gois, "A review on Generative Adversarial Networks for image generation," *Comput Graph*, vol. 114, pp. 13–25, Aug. 2023, doi: https://doi.org/10.1016/j.cag.2023.05.010.

[26] X. Zhao, T. Yang, B. Li, and X. Zhang, "SwinGAN: A dual-domain Swin Transformer-based generative adversarial network for MRI reconstruction," *Comput. Biol. Med.*, vol. 153, pp. 106513–106513, Feb. 2023, doi: https://doi.org/10.1016/j.compbiomed.2022.106513.

[27] Q. Zhang, Y. Zheng, Q. Yuan, M. Song, H. Yu, and Y. Xiao, "Hyperspectral Image Denoising: From Model-Driven, Data-Driven, to Model-Data-Driven," *IEEE Trans. Neural Networks Learn. Syst.*, pp. 1–21, Jan. 2023, doi: https://doi.org/10.1109/tnnls.2023.3278866.

[28] Z. Wang, L. Zhao, T. Zhong, Y. Jia, and Y. Cui, "Generative adversarial networks with multi-scale and attention mechanisms for underwater image enhancement," *Front. Mar. Sci.*, vol. 10, Oct. 2023, doi: https://doi.org/10.3389/fmars.2023.1226024.

[29] Z. Tang, C. Wu, Y. Xiao, and C. Zhang, "Evaluation of Painting Artistic Style Transfer Based on Generative Adversarial Network," Apr. 2023, doi: https://doi.org/10.1109/icccbda56900.2023.10154714.

[30] X. Lu, X. Xie, C. Ye, H. Xing, Z. Liu, and C. Cai, "A lightweight generative adversarial network for single image super-resolution," *The Visual Computer*, Feb. 2023, doi: https://doi.org/10.1007/s00371-022-02764-z.

[31] H. Hou, J. Xu, Y. Hou, X. Hu, B. Wei, and D. Shen, "Semi-Cycled Generative Adversarial Networks for Real-World Face Super-Resolution," *IEEE Trans. Image Process.*, vol. 32, pp. 1184–1199, Jan. 2023, doi: https://doi.org/10.1109/tip.2023.3240845.

[32] Y. Chen, D. Ji, Q. Ma, C. Zhai, and Y. Ma, "A Novel Generative Adversarial Network for the Removal of Noise and Baseline Drift in Seismic Signals," *IEEE Trans. Geosci. Remote Sens.*, pp. 1–1, Jan. 2024, doi: https://doi.org/10.1109/tgrs.2024.3358901.

[33] X. Liu, F. Lyu, L. Chen, C. Li, S. Zu, and B. Wang, "Seismic random noise suppression based on deep image prior and total variation," *IEEE Trans. Geosci. Remote Sens.*, pp. 1–1, Jan. 2024, doi: https://doi.org/10.1109/tgrs.2024.3371714.

[34] M. Zhao, X. Pan, S. Xiao, Y. Zhang, C. Tang, and X. Wen, "Seismic Data Interpolation Based on Spectrally Normalized Generative Adversarial Network," *IEEE Trans. Geosci. Remote Sens.*, vol. 61, pp. 1–11, Jan. 2023, doi: https://doi.org/10.1109/tgrs.2023.3301270.

[35] M. Ding, Y. Zhou, and Y. Chi, "Self-Attention Generative Adversarial Network Interpolating and Denoising Seismic Signals Simultaneously," *Remote Sens.*, vol. 16, no. 2, pp. 305–305, Jan. 2024, doi: https://doi.org/10.3390/rs16020305.

[36] S. Sun, L. Zhao, H. Chen, Z. He, and J. Geng, "Pre-stack seismic inversion for elastic parameters using model-data-driven generative adversarial networks," *Geophysics.*, pp. 1–91, Dec. 2022, doi: https://doi.org/10.1190/geo2022-0314.1.

[37] F. Yang and J. Ma, "FWIGAN: Full-Waveform Inversion via a Physics-Informed Generative Adversarial Network," *J GEOPHYS RES-SOL EA.*, vol. 128, no. 4, Apr. 2023, doi: https://doi.org/10.1029/2022jb025493.

[38] L. Lin, Z. Zhong, C. Cai, C. Li, and H. Zhang, "SeisGAN: Improving Seismic Image Resolution and Reducing Random Noise Using a Generative Adversarial Network," *Math. Geosci.*, Oct. 2023, doi: https://doi.org/10.1007/s11004-023-10103-8.

[39] Q.-F. Sun, J.-Y. Xu, H.-X. Zhang, Y.-X. Duan, and Y.-K. Sun, "Random noise suppression and super-resolution reconstruction algorithm of seismic profile based on GAN," *J. Pet. Explor. Prod. Technol.*, Jan. 2022, doi: https://doi.org/10.1007/s13202-021-01447-0.

[40] N. D. Lynn, A. I. Sourav, and A. J. Santoso, "Implementation of Real-Time Edge Detection Using Canny and Sobel Algorithms," *IOP Conf. Ser.: Mater. Sci. Eng.*, vol. 1096, no. 1, p. 012079, Mar. 2021, doi: https://doi.org/10.1088/1757-899x/1096/1/012079.

[41] H. R. Sheikh, M. F. Sabir and A. C. Bovik, "A Statistical Evaluation of Recent Full Reference Image Quality Assessment Algorithms," *IEEE Trans. Image Process.*, vol. 15, no. 11, pp. 3440-3451, 2006. https://doi.org/10.1109/TIP.2006.881959.

[42] D. R. I. M. Setiadi, "PSNR vs SSIM: imperceptibility quality assessment for image steganography," *Multimed Tools Appl*, Nov. 2020, doi: https://doi.org/10.1007/s11042-020-10035-z.